\documentstyle[12pt]{article}
\begin{document}

\def\0{\c{s}}
\def\1{\"{o}}
\def\beq{\begin{equation}}
\def\eeq{\end{equation}}
\def\bea{\begin{eqnarray}}
\def\eea{\end{eqnarray}}
\def\ve{\vert}
\def\vel{\left|}
\def\ver{\right|}
\def\nnb{\nonumber}
\def\ga{\left(}
\def\dr{\right)}
\def\aga{\left\{}
\def\adr{\right\}}
\def\rar{\rightarrow}
\def\nnb{\nonumber}
\def\la{\langle}
\def\ra{\rangle}
\def\lla{\left<}
\def\rra{\right>}
\def\ba{\begin{array}}
\def\ea{\end{array}}
\def\ds{\displaystyle}


\title{ {\Large {\bf 
Short range correlations in a one dimensional electron gas } } }

\author{\vspace{1cm}\\
{\small Murat Tas \thanks 
{e-mail: tasm@metu.edu.tr}\,\, and
Mehmet Tomak \thanks 
{e-mail: tomak@metu.edu.tr}} \\
{\small Physics Department, Middle East Technical University} \\
{\small 06531 Ankara, Turkey} }

\date{}

\begin{titlepage}
\maketitle
\thispagestyle{empty}

\begin{abstract}
We use the SSTL (Singwi, Sj\1lander, Tosi, Land) approximation to
investigate the short--range correlations in a one dimensional 
electron gas, for the first time. Although SSTL is introduced to 
better satisfy the compressibility sum rule in three dimensions, 
the widely used STLS (Singwi, Tosi, Land, Sj\1lander) approximation 
turns out to be more successful in the case of the one dimensional
electron gas.
\end{abstract}

~~~Keywords: Quantum wires; Screening; Compressibility
\end{titlepage}

\section{Introduction}
The advances in fabrication technologies have made it possible to 
experimentally fabricate one dimensional electronic
structures\cite{1}--\cite{5}. This has naturally resulted in an 
increasing interest in theoretical investigations of such
structures\cite{6}--\cite{13}. The resulting one dimensional electron 
gas has also been subject to computational investigation\cite{14}. 

The extensive investigations in the three dimensional electron gas have 
clearly shown the importance of the short range correlations at lower 
densities in determining the pair distribution function $g(r)$ at small 
$r$\cite{15}.

In the high density limit, the long range correlations can be described
well by the random phase approximation (RPA). One way to include the 
short range correlation effects beyond RPA is to use the powerful 
approach developed by STLS\cite{16}. 

It is well known that STLS approach suffers from a compressibility
inconsistency. The compressibility calculated from the small $q$ limit 
of the dielectric function $\varepsilon(q)$ does not agree with that 
calculated from the ground state energy. There have been several 
attempts to overcome this inconsistency\cite{17,18}. SSTL\cite{17} 
included the effect of screening on the effective interaction potential 
as explained in the next section.

In this work, we aim to check the performance of the SSTL approach and 
compare it with the STLS in a one dimensional electron gas. To the best 
of our knowledge, this is the first application of the SSTL approach 
in lower dimensions.

The organization of the paper is as follows; the STLS and SSTL 
formalisms are given in section 2. The results and discussion 
concentrating on the comparison between STLS and SSTL performance
on the compressibility issue are presented in section 3.

\section{Formalism}
In the mean field approximation, a key component of the electron gas 
is the usual static structure factor S(q) which is related to 
density--density response function $\chi(q,\omega)$ through the 
fluctuation--dissipation theorem as

\begin{equation}
S(q)=-\frac{1}{n \pi}\int_0^{\infty} d\omega~ Im\chi(q,\omega).
\end{equation}
The density--density response function $\chi(q,\omega)$ is defined as

\beq
\chi(q,\omega)=\frac{\chi_0(q,\omega)}{1-V_{eff}(q)\chi_0(q,\omega)},
\eeq
where $\chi_0(q,\omega)$ is the zero--temperature susceptibility of
a noninteracting electron gas and is given in one dimension by

\beq
\chi_0(q,\omega)=\chi_{01}(q,\omega)+i \chi_{02}(q,\omega),
\eeq
with

\beq
\chi_{01}(q,\omega)=\frac{m^{\star}}{\hbar^2\pi q}\ln\left|
\frac{\omega^2-\omega_{-}^2}{\omega^2-\omega_{+}^2}\right|,
\eeq
and
\[\chi_{02}(q,\omega)=\left\{ \begin{array}{ll}
-{\ds \frac{m^{\star}}{\hbar^2 q}}, & \mbox
{~~~~$\omega_{-} < \omega < \omega_{+}$} \\
\\
0 &~~~~ \mbox{otherwise,} \end{array}\right. \] 
where ${\ds \omega_{\pm}=\left| \frac{\hbar^2 q^2}{2m^{\star}}
\pm \frac{\hbar q k_F}{m^{\star}} \right|}$ are the boundaries for 
particle--hole excitations. The Fermi wave vector $k_F$ is related 
to the one dimensional electron density $n$ via ${\ds k_F=n \pi/2}$. 
The dimensionless electron density parameter is defined as 
${\ds r_s=\pi/(4k_F a_B^{\star})}$, where ${\ds a_B^{\star}}$ is 
the effective Bohr radius. 

$V_{eff}(q)$ in Eq. (2) is the self--consistent effective potential
related to the static structure factor through

\beq
V_{eff}(q)=v(q)+\frac{1}{n q}\int_0^{\infty} \frac{dk}{2\pi} 
\left[S(k)-1\right]\left\{(q+k)v(q+k)+(q-k)v(q-k)\right\},
\eeq
where $v(q)$ is the Fourier transform of the Coulomb interaction 
between two electrons in the lowest subband in the harmonic 
confinement in one dimension and is given by 

\beq
v(q)=\frac{2e^2}{\varepsilon_0}~F(q),
\eeq
where $\varepsilon_0$ is the background dielectric constant, $q$ 
is the wave vector along the wire and $F(q)$ is the form factor 
which takes into account the finite thickness of the wire. The 
form factor reads

\beq
F(q)=\frac{1}{2}~exp\left(\frac{b^2 q^2}{4}\right)
K_0 \left(\frac{b^2 q^2}{4}\right),
\eeq
here $K_0(x)$ is the zeroth order modified Bessel function of the
second kind and $b$ is the characteristic length of the harmonic
potential. Indeed, it is a measure of the effective radius of the 
quantum wire\cite{Hu}.

In the STLS approximation,

\beq
\chi(q,\omega)=\frac{\chi_0(q,\omega)}{1-v(q)[1-G(q)]\chi_0(q,\omega)},
\eeq
where the local field correction $G(q)$ is

\beq
G(q)=-\frac{1}{n}\int_{-\infty}^{\infty} ~\frac{dk}{2\pi}~
\frac{k}{q}~\frac{v(k)}{v(q)}~[S(q-k)-1].
\eeq
The set of Eqs. (1), (8) and (9) have to be solved self--consistently 
for $G(q)$, $\chi(q)$ and $S(q)$ within the STLS approximation. In RPA,
$G(q)=0$.

The SSTL approximation is different from the STLS approximation in 
that the potential under the integral sign in Eq. (5) is screened by 
the static dielectric function $\varepsilon(q)$ which is given by,

\beq
\varepsilon(q)=1-\frac{v(q)\chi_0(q)}{1+G(q)v(q)\chi_0(q)}.
\eeq
This is originally done to better satisfy the compressibility sum 
rule in three dimensional electron gas.

The ground state energy per particle in one dimensional electron 
gas may be written as\cite{Vera}

\beq
\varepsilon_g=\varepsilon_{kin}+\varepsilon_{ex}+\varepsilon_{cor},
\eeq 
where ${\ds \varepsilon_{kin}}$ is the kinetic energy per particle,
and simply ${\ds \varepsilon_{kin}=\pi^2/(48r_s^2)}$ in units of
effective Rydberg $Ry^{\star}$. ${\ds \varepsilon_{ex}}$ is the exchange
energy per particle

\beq
\varepsilon_{ex}=-\frac{1}{2\pi^2 n}\int_0^{k_F}~dq~ 
\int_{-(k_F+q)}^{k_F-q} dk~v(k),
\eeq

and ${\ds \varepsilon_{cor}}$ is the correlation energy per particle

\beq
\varepsilon_{cor}=\frac{1}{2\pi r_s}\int_0^{r_s}~dr_s^{\prime}~
\int_0^{\infty}~dq~v(q)~\left[S(q,r_s^{\prime})-1\right].
\eeq

The compressibility $K$ may be calculated either by using the ground 
state energy per particle $\varepsilon_g$

\beq
\frac{1}{K}=n^2~\frac{d^2}{dn^2}~(n~\varepsilon_g),
\eeq 
or by using the $q \rightarrow 0$ limit of the static dielectric 
function $\varepsilon(q)$

\beq
\lim_{q \rightarrow 0}~\varepsilon(q)=1+v(q)~n^2~K.
\eeq

The compressibility sum rule is that the compressibilities calculated 
by using Eq. (14) and Eq. (15) are the same.

\section{Results and Discussion}
In Fig. 1 the local field correction $G(q)$ calculated 
self--consistently using STLS and SSTL approximations are shown for 
fixed ${\ds b=5a_B^{\star}}$ and $r_s=5$. The most striking difference 
between the two curves is their large $q$ limits. As STLS $G(q)$ is 
approaching 1, SSTL $G(q)$ approaches a limit bigger than 1. It may 
be noted that the relation\cite{18} $G(\infty)=1-g(0)$ is satisfied 
in our calculations. This, of course, is going to lead to different 
results when applied to physical properties of the one dimensional
electron gas. 

The different large $q$ behaviour of $G(q)$ is expected to lead to 
different small $r$ behaviour of the pair correlation function $g(r)$. 
This is shown in Fig. 2. As radius of the wire gets smaller the 
difference between STLS and SSTL results become larger, as may be seen 
in Fig. 3. As the density becomes smaller (i.e., $r_s$ becomes larger) 
the SSTL $g(r)$ becomes negative for small $r$. This limit is where the 
correlation effects become important.

The SSTL ${\ds \varepsilon(q)^{-1}}$ is given in Fig. 4. This is rather 
similar to STLS or RPA ${\ds \varepsilon(q)^{-1}}$ with a discontinuity 
in the derivative at ${\ds q=2k_F}$. 

The ground state energy per particle for our system is shown in Fig. 5. 
The STLS and SSTL curves are rather similar.

The compressibility calculated by using the ground state energy in 
three different approaches is presented in Fig. 6 as a function of 
$r_s$. The similarity in energies is also reflected in these curves. 
The negative compressibility values at higher $r_s$ show that the 
system becomes unstable.

The compressibility calculated by using $q \rightarrow 0$ limit of the 
static dielectric function is shown in Fig. 7. Here, the STLS and SSTL 
compressibilities calculated by two routes are as different as in RPA. 
This surprising result is a natural consequence of the small $q$ 
behaviour of the local field correction which is very small for SSTL 
at small $q$ values. 

The compressibility in a one dimensional system is studied previously 
by Gold and Calmels\cite{7} using a three-sum-rule approach for the 
local field correction. The compressibility sum rule is better satisfied
within this variant of STLS in one dimension as can be seen from their
figure 4. We find the same result by the present full STLS calculation.

We can conclude that the short--range correlations are calculated using 
the SSTL approximation in a one dimensional electron gas, for the first
time. The performance of the SSTL approximation is compared with the 
more widely used STLS approximation. It is shown that SSTL compares 
reasonably well with STLS when the pair correlation function, dielectric 
function and the ground state energy per particle are considered but 
fails totally when the compressibility sum rule is checked.

\newpage
\begin{figure}
\vskip 1cm
    \includegraphics{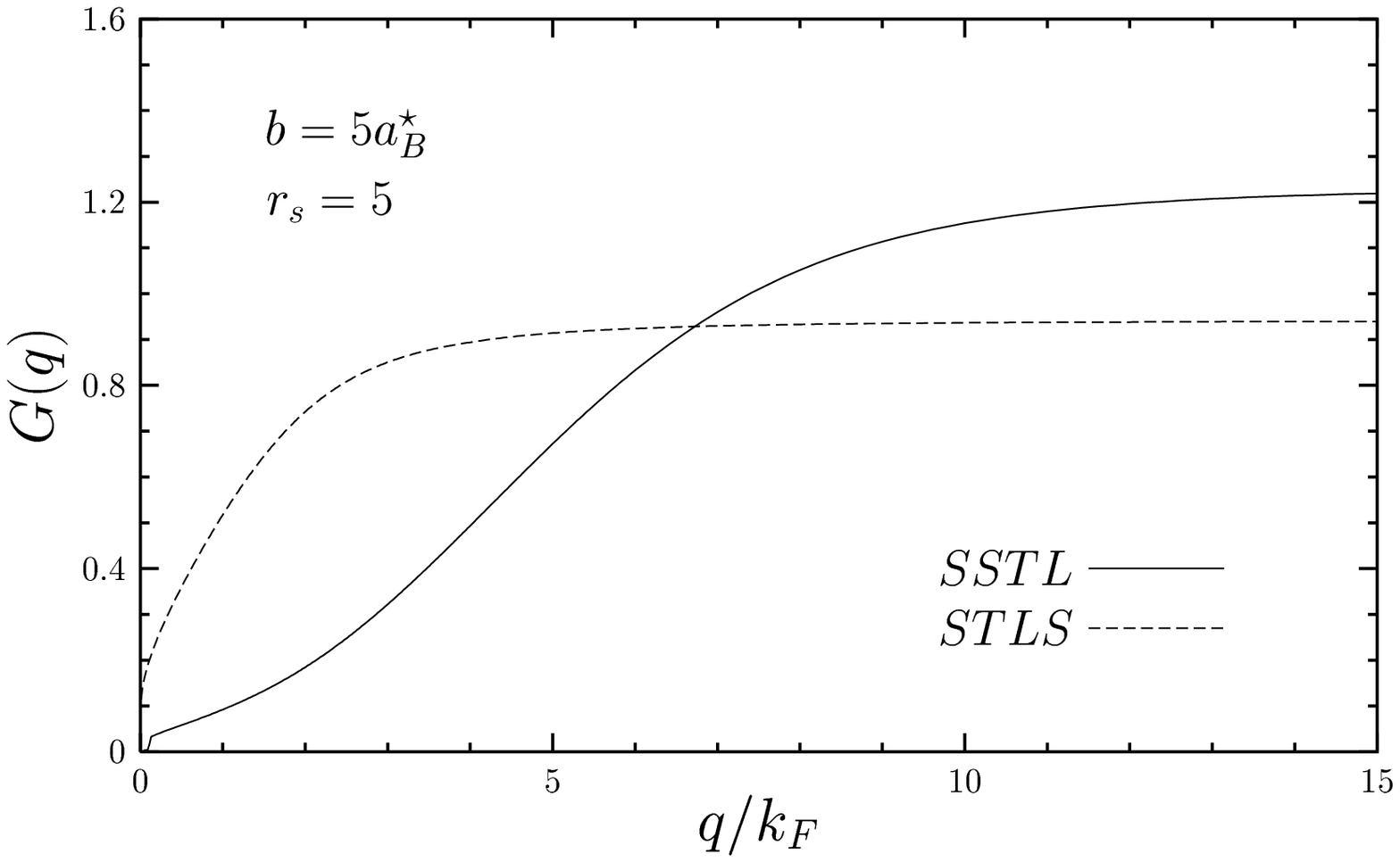}
\vskip 8.1cm
\caption{The local field correction in SSTL and STLS approximations
for wire radius ${\ds b=5a_B^{\star}}$ and ${\ds r_s=5}$.}
\end{figure}

\begin{figure}
\vskip 1cm
    \includegraphics{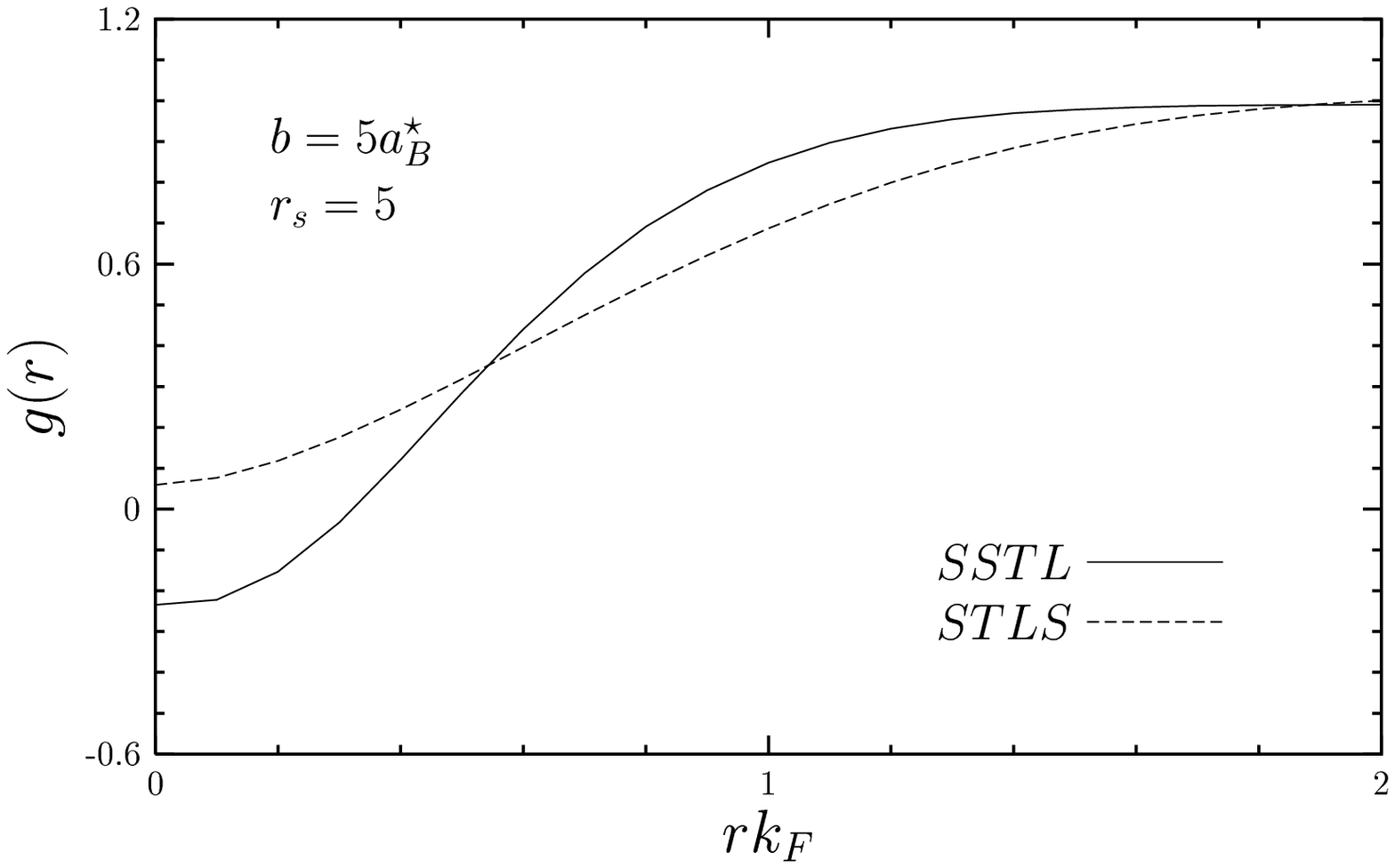}
\vskip 8.1cm
\caption{The pair correlation function in SSTL and STLS approximations 
for wire radius ${\ds b=5a_B^{\star}}$ and ${\ds r_s=5}$.}
\end{figure}

\begin{figure}
\vskip 1cm
    \includegraphics{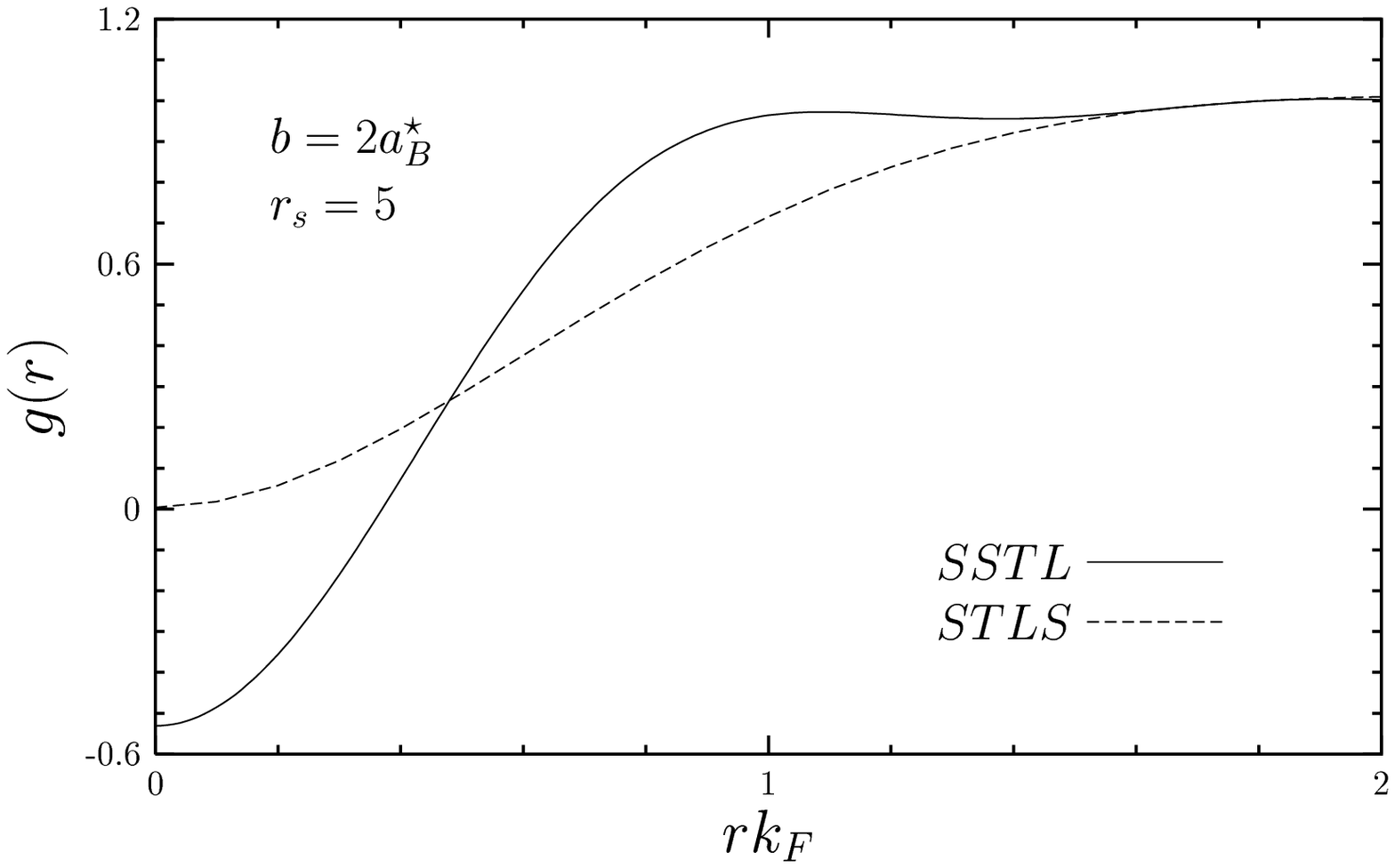}
\vskip 8.1cm
\caption{The pair correlation function in SSTL and STLS approximations
for wire radius ${\ds b=2a_B^{\star}}$ and ${\ds r_s=5}$.}
\end{figure}

\begin{figure}
\vskip 1cm
    \includegraphics{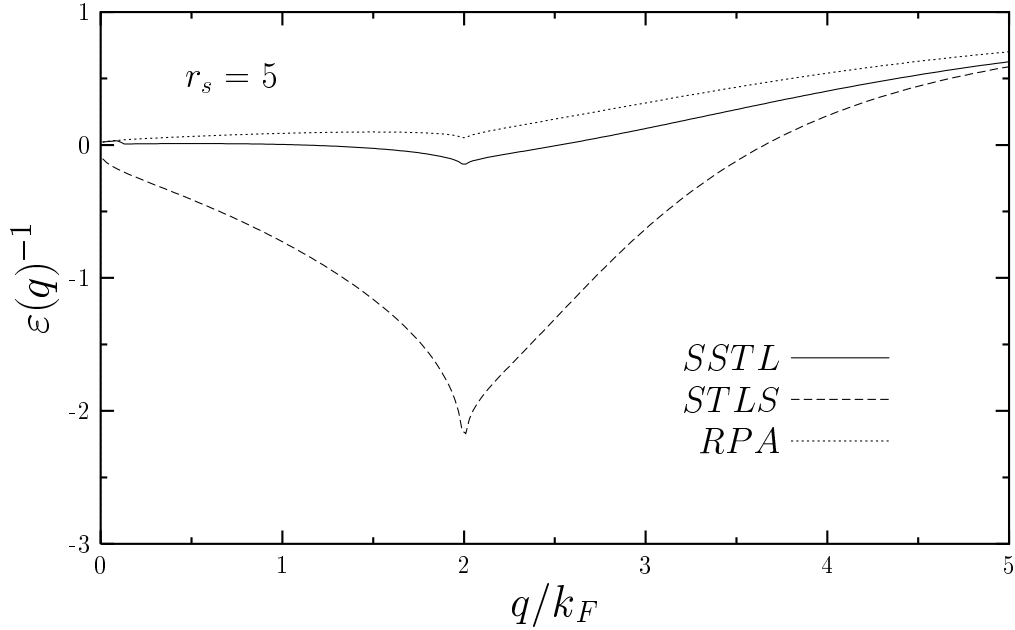}
\vskip 8.1cm
\caption{The inverse dielectric function in different approximations 
for wire radius ${\ds b=5a_B^{\star}}$ and ${\ds r_s=5}$.}
\end{figure}

\begin{figure}
\vskip 1cm
    \includegraphics{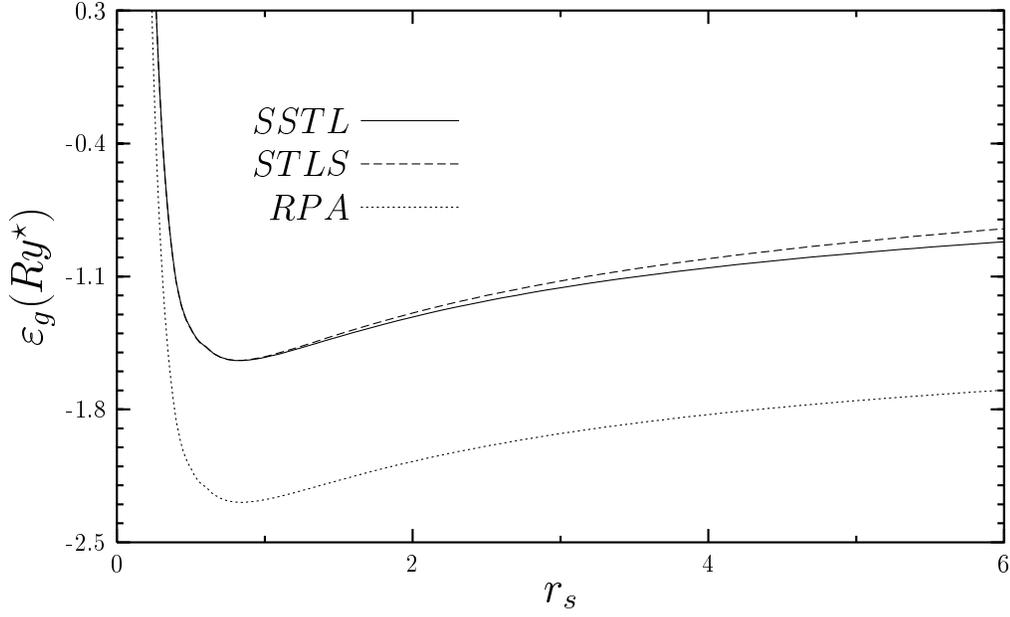}
\vskip 8.1cm
\caption{The ground state energy per particle in different approximations
for wire radius ${\ds b=2a_B^{\star}}$.}
\end{figure}

\begin{figure}
\vskip 1cm
    \includegraphics{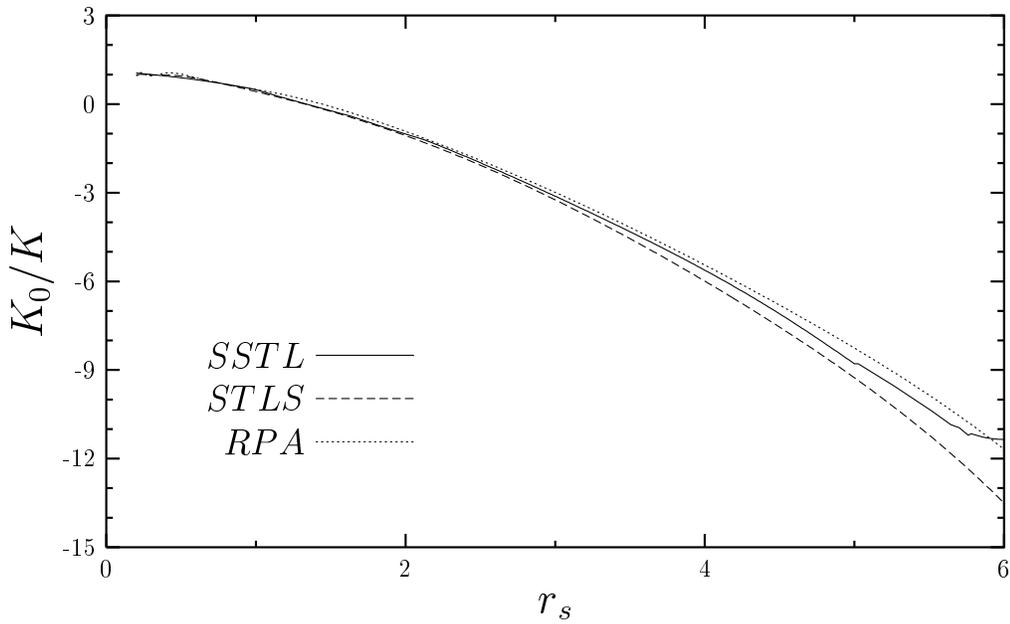}
\vskip 8.1cm
\caption{The compressibility calculated by using the ground state energy
in three different approaches for wire radius ${\ds b=2a_B^{\star}}$. 
Hereafter, $K_0$ is the free electron compressibility.}
\end{figure}

\begin{figure}
\vskip 1cm
    \includegraphics{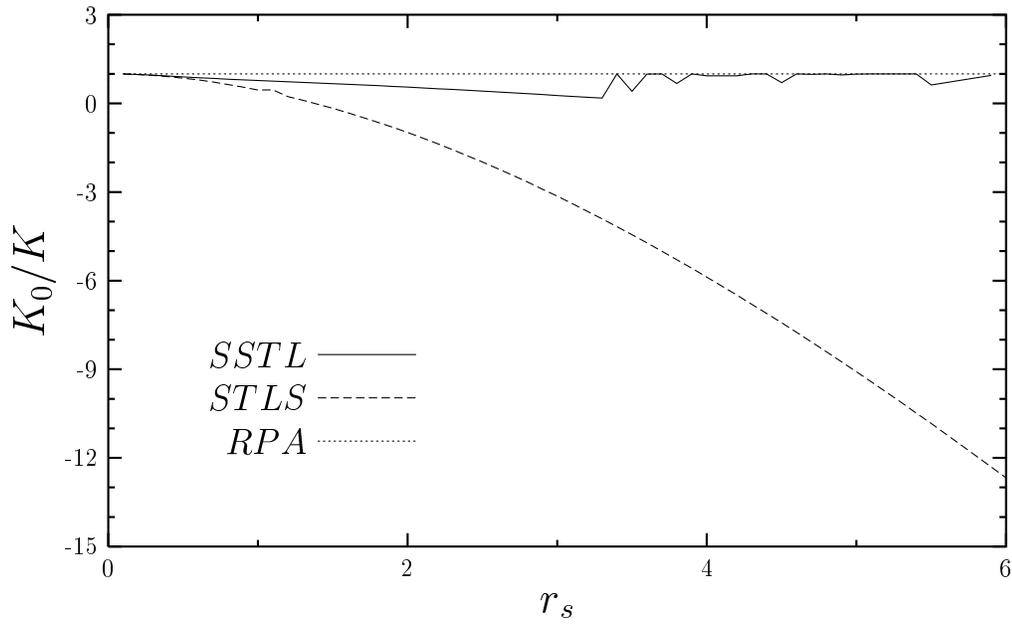}
\vskip 8.1cm
\caption{The compressibility calculated by using the 
${\ds \lim_{q \rightarrow 0}~\varepsilon(q,0)}$ in three different
approaches for wire radius ${\ds b=2a_B^{\star}}$.}
\end{figure}
\end{document}